# Broadband dielectric spectroscopy on benzophenone: α relaxation, β relaxation, and mode coupling theory

P. Lunkenheimer[1],* L. C. Pardo[1,2,+], M. Köhler[1], and A. Loidl[1]

[1]Experimental Physics V, Center for Electronic Correlations and Magnetism, University of Augsburg, 86135 Augsburg, Germany

[2]Departament de Física i Enginyeria Nuclear, ETSEIB, Universitat Politècnica de Catalunya, Diagonal 647, 08028 Barcelona, Catalonia, Spain

We have performed a detailed dielectric investigation of the relaxational dynamics of glass-forming benzophenone. Our measurements cover a broad frequency range of 0.1 Hz to 120 GHz and temperatures from far below the glass temperature well up into the region of the small-viscosity liquid. With respect to the α relaxation this material can be characterized as a typical molecular glass former with rather high fragility. A good agreement of the α relaxation behavior with the predictions of the mode coupling theory of the glass transition is stated. In addition, at temperatures below and in the vicinity of $T_g$ we detect a well-pronounced β relaxation of Johari-Goldstein type, which with increasing temperature develops into an excess wing. We compare our results to literature data from optical Kerr effect and depolarized light scattering experiments, where an excess-wing like feature was observed in the 1 – 100 GHz region. We address the question if the Cole-Cole peak, which was invoked to describe the optical Kerr effect data within the framework of the mode coupling theory, has any relation to the canonical β relaxation detected by dielectric spectroscopy.

## I. INTRODUCTION

The dynamics of glassy matter reveals a rich variety of unusual and so far only poorly understood phenomena [1,2]. Thus the investigation of glasses and supercooled liquids with spectroscopic methods is a lively field, which in recent years gained additional momentum due to new experimental developments allowing for the investigation of increasingly broader time and frequency ranges. But also the theoretical development kept pace with the experiment: Various models appeared that not only account for the well-known α relaxation, characterizing the slowing down dynamics when approaching the glass transition, but also predict other, faster processes (e.g., [3,4,5,6,7,8,9]). One of the most prominent experimentally detected fast processes is the so-called Johari-Goldstein (JG) β relaxation [10]. In spectra of the dielectric loss $\varepsilon''$, it leads to a second peak in addition to that related to the α relaxation. Depending on temperature, it is mostly observed in the Hz - MHz range, at frequencies beyond that of the α peak. As shown by Johari and Goldstein [10] it is inherent to the glassy state of matter. Some glass formers do not show a β relaxation, but a so-called excess wing, a second, shallower power-law decrease at the high-frequency flank of the α peak. This feature is known since long [11,12,13,14,15], but only recently was discovered to be due to a β relaxation peak [16,17,18,19].

The most prominent theory for the description of glassy dynamics and the glass transition is the mode coupling theory (MCT) [3]. It not only describes the typical features of the α relaxation, but also predicts a so-called fast β relaxation, showing up in the GHz - THz range. This fast process is ascribed to the "rattling" movement of a particle in the transient cage formed by its neighbors, considered stable on a short time scale. It should be noted that the fast β process is not identical with the JG β relaxation. To distinguish the latter and other secondary relaxations in the Hz – MHz range from the fast β process of MCT, they often are termed "slow β processes". In the basic versions of MCT, a combination of asymptotic power laws is predicted to describe the spectra in the regime of the fast process. They do not form a peak but define the low and high-frequency wing of a shallow minimum in the frequency-dependent loss, which is followed by the so-called microscopic peak located in the THz range. Both exponents of the asymptotic power laws are determined by a system parameter $\lambda$. While the validity of MCT often is quite controversially discussed and there are many competing approaches (e.g., [5,6,7,20]), some consensus seems to have developed in the glass community that MCT provides a correct description of glassy dynamics, at least for high temperatures, $T > T_c$. The critical temperature $T_c$ can be regarded as idealized glass transition temperature often roughly located at 1.2 $T_g$ with $T_g$ the experimental glass temperature. The main success of MCT is the prediction of the fast $\beta$ relaxation, which was experimentally verified in numerous investigations (for reviews, see [21]). However, the original MCT only revealed three time scales of dynamic processes, that of the $\alpha$ relaxation, the fast $\beta$ process and the microscopic process. Thus it seemed that MCT cannot describe the JG $\beta$ relaxation or excess wings showing up in most glass formers when approaching $T_g$ and below. Instead in the simplest versions of MCT, the high-frequency flank of the $\alpha$ relaxation peak directly crosses over into the low-frequency wing of the minimum, the so-called von-Schweidler law $\nu^b$. At high temperatures this limitation of MCT represents no problem: There the JG $\beta$ peak or the excess wing usually is merged with the $\alpha$ peak and no longer plays any role in the spectra [2,12,13,14,15]. However, at lower temperatures where excess wing or $\beta$ peak develops, MCT must fail in a complete description of the spectra.

*Corresponding author. peter.lunkenheimer@physik.uni-augsburg.de
+Present address: Forschungszentrum Jülich GmbH, 52425 Jülich, Germany

Recently it was shown that basic MCT could not describe results from optical Kerr effect (OKE) experiments on benzophenone and several other glass formers [22,23], even at rather high temperatures, $T > T_c$. In the OKE response function, defined in time domain, a power law decay $t^{a-1}$ occurred, with $a$ in the region $0 - 0.21$. In frequency domain, $a = 0$ would correspond to nearly constant loss in the susceptibility and $a > 0$ leads to $\chi'' \sim \nu^a$. These data recently were analyzed by Götze and Sperl [24,25] by taking into account rotation-translational coupling in a schematic MCT model. Based on earlier theoretical work [26], they demonstrated that under certain circumstances instead of a minimum the fast $\beta$ relaxation of MCT also can lead to a peak or at least an excess wing. It can be approximately described by the empirical Cole-Cole (CC) law [27]:

$$\varepsilon^* = \varepsilon_{\infty,CC} + \frac{\Delta\varepsilon_{CC}}{1+(i\omega\tau_{CC})^{1-\alpha_{CC}}} \quad (1)$$

($\varepsilon^*$ is the complex dielectric permittivity, $\varepsilon_{\infty,CC}$ the high-frequency limiting value of the real part, $\Delta\varepsilon_{CC}$ the relaxation strength, $\omega = 2\pi\nu$ the circular frequency, $\tau_{CC}$ the relaxation time, and $\alpha_{CC}$ the width parameter [28]). The experimental results in BZP indeed could be fitted assuming that the CC peak position is much lower than that of the microscopic peak so that the low-frequency wing of the CC peak is superimposed by the $\alpha$ peak [25]. Then the high-frequency wing of the CC peak should show up in the spectrum as a second more shallow power law at the high-frequency flank of the $\alpha$ peak (a schematic view of this scenario is provided in Fig. 1(b) of Ref. 29). In time domain this leads to a power law just as experimentally observed [22,23]. Very recently depolarized light scattering (DLS) experiments on BZP were reported [30]. In accordance with expectation (for both techniques the same correlation function applies), the results were fully consistent with the ones from OKE. In that work also loss spectra were presented obtained by Fourier transformation of the OKE [23] and DLS [30] response functions. As expected, a second more shallow power law was found for the lowest investigated temperatures.

Interestingly, it is common practice to use the CC function also for the description of the JG $\beta$ relaxation. Indeed it was speculated [25,31] that the new CC peak of MCT could be related to the JG process. This was also motivated by the fact that the spectrum at 251 K in BZP looks very similar as typical spectra with excess wing found in various glass formers [30,32]. However, one should be aware that usually excess wings and $\beta$ relaxations are reported at temperatures far below $T_c$ and at high temperatures these features merge with the $\alpha$ peak. The excess wing becomes most pronounced at low temperatures and frequencies [2,12,14] and even can develop into a JG $\beta$ peak when approaching $T_g$ [16,17]. Therefore it is suggestive to investigate BZP at lower frequencies and temperatures. However, due to experimental limitations, the lowest frequencies covered by OKE and DLS are about 10-100 MHz. In contrast, in dielectric spectroscopy, while high frequencies are difficult to access, lower frequencies down to the Hz frequency range or lower are routinely investigated. Indeed the JG-relaxation and excess wing phenomena were discovered and most intensively investigated by dielectric spectroscopy [2,10,12,13,14,15,16,33]. It is known since long that BZP (diphenylketone, $H_5C_6-CO-C_6H_5$) can be supercooled [34] and a glass temperature of $T_g = 212$ K was reported [35]. However, to our knowledge so far there is no dielectric investigation of this glass former. Thus, in [29] we have provided broadband dielectric spectra on BZP covering frequencies from 0.1 Hz to 30 GHz at temperatures extending well below $T_g$. In that work we have focused especially on the $\beta$ relaxation and excess wing and their relation to the feature seen in the light scattering data. In the present work we provide a more thorough discussion of the relaxational properties on BZP, also providing a detailed characterization of the $\alpha$ relaxation of this glass former with information on fragility, stretching, and relaxation strength. In addition, first dielectric data from quasi-optic measurements at 60 - 120 GHz are reported.

## II. EXPERIMENTAL DETAILS

Benzophenone was purchased from Merck with a minimum purity of 99% and used without further purification. Additional measurements in distilled material revealed identical results. For the dielectric measurements up to 3 GHz, the sample material was heated to the liquid state (melting point $\approx 322$ K) and filled into pre-heated home-made parallel-plate capacitors [36]. Dielectric spectra in the frequency region up to 3 MHz were obtained by means of frequency response analysis using a Novocontrol $\alpha$-analyzer. At frequencies $1$ MHz $\leq \nu \leq 3$ GHz the impedance analyzer

Agilent E4991A was employed using a reflectometric technique, with the sample capacitor mounted at the end of a 7 mm coaxial line [36,37]. In the region 300 MHz ≤ $\nu$ ≤ 30 GHz an Agilent E8363 network analyzer was used to measure the reflection of a coaxial line whose end was immersed into the sample material [38]. Quasi-optic measurements at frequencies of 60 - 120 GHz were carried out using a Mach-Zehnder interferometer. It allows measuring the frequency dependence of both, the transmission and the phase shift, of a monochromatic electromagnetic beam through the sample [36,39]. To hold the sample material, commercially available quartz cuvettes were used. The data were analyzed using standard optical formulae for multilayer interference with the known thickness and optical parameters of the cuvette windows in order to get the complex dielectric permittivity of the sample.

In the region of 230–255 K, BZP exhibits an enhanced crystallization tendency. Dielectric spectra at $T \leq 235$ K were collected under heating after passing this region with rapid cooling rates of up to 10 K/min. Spectra between 240 K and 250 K were measured during quickly cooling the sample. Rates up to 3 K/min were necessary to avoid crystallization. Thus in this region the thermal equilibration of the sample may not have been perfect and a temperature error up to 1 K has to be assumed. Due to experimental problems arising from the strong crystallization tendency and the relatively low loss of BZP, the quasi-optic measurements could be performed at high temperatures only.

## III. RESULTS AND DISCUSSION

### A. $\alpha$ relaxation

Figure 1 shows the frequency dependence of the dielectric loss of BZP for various temperatures. The spectra are dominated by well-pronounced peaks strongly shifting to lower frequencies with decreasing temperature. They can be ascribed to the $\alpha$ relaxation, their continuous shift reflecting the glassy freezing of molecular dynamics when $T_g$ is approached. In addition, an excess wing (e.g., at 225 K and $\nu > 1$ MHz) and a $\beta$ relaxation (e.g. at 200 K and $\nu > 100$ Hz) are detected, which both will be treated in the next section. The $\alpha$ relaxation in BZP so far only was investigated at temperatures down to 250 K and frequencies down to 30 MHz by OKE and DLS [22,23,30]. The present spectra extend this region down to temperatures significantly below $T_g \approx 212$ K and down to frequencies of 0.1 Hz. To obtain information on the characteristic parameters of the relaxation, least square fits of the experimental data were performed. The empirical Cole-Davidson (CD) function [40],

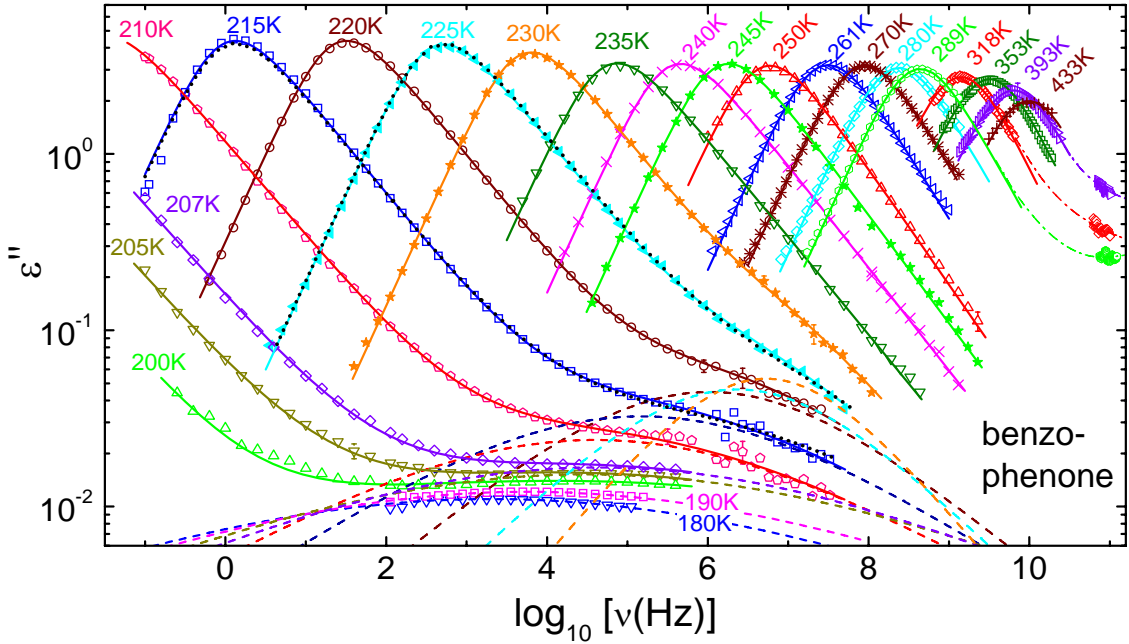

FIG. 1. (Color) Frequency-dependent dielectric loss of BZP for various temperatures. The solid lines are fits with the sum of a CD [eq. (2)] and a CC function [eq. (1)] (at $T \leq 190$ K the CD amplitude and at $T \geq 235$ K the CC amplitude were set to zero). The dashed lines show the CC part of the fits accounting for the observed $\beta$ relaxation. The dotted lines represents fits of the 215 and 225 K curves performed with the convolution ansatz promoted in [62]. The dash-dotted lines at high frequencies are drawn to guide the eyes.

$$\varepsilon^* = \varepsilon_{\infty,CD} + \frac{\Delta\varepsilon_{CD}}{(1+i2\pi\nu\tau_{CD})^{\beta_{CD}}} \qquad (2)$$

was found to provide reasonable fits of the spectra (solid lines in Fig. 1). At temperatures $T \leq 230$ K, a Cole-Cole (CC) function, eq. (1), was added to the fitting function to account for the excess wing and $\beta$ relaxation as detailed in the next section.

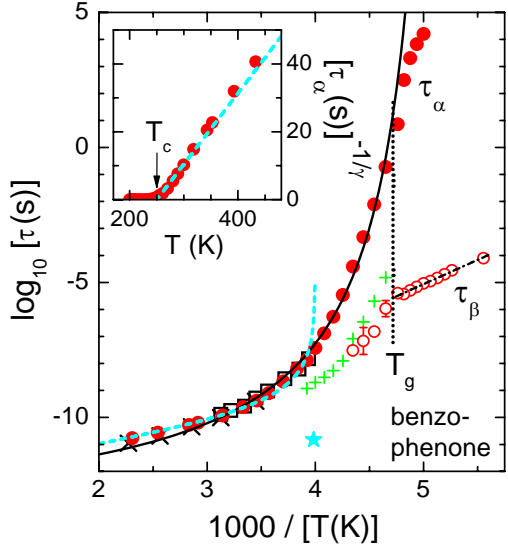

FIG. 2. (Color) Relaxation times determined from the fits shown in Fig. 1 (full and open circles) [29]. In addition, literature data from OKE [23] and DLS [30] experiments are shown (open squares and ×'s, respectively). The solid line is a fit of the dielectric data with a VFT function, eq. (3) ($\tau_0 = 4.2\times10^{-13}$ s, $T_{VF} = 189$ K, and $D = 3.8$). The dashed line is a fit of the dielectric data at $T \geq 255$ K with eq. (4), using the parameters $T_c = 250$ K and $\gamma = 1.92$ as reported by Cang et al. [23]. The dash-dotted line is a fit of the $\beta$ relaxation time at $T < T_g$ assuming an Arrhenius law ($\tau_0 = 8.8\times10^{-15}$ s, $E = 0.36$ eV). The pluses show the primitive relaxation times of CM, calculated from eq. (5) [67]. The star shows the CC relaxation time as determined by Sperl [25] from a MCT analysis of the OKE data. The inset shows the dielectric data and MCT fit, using a representation that should linearize according to the MCT prediction, eq. (4), with $\gamma = 1.92$ [23].

The temperature dependence of the $\alpha$ relaxation time $\tau_\alpha = \tau_{CD}$ resulting from these fits is shown in Fig. 2 (closed circles) [29]. Obviously, $\tau_\alpha(T)$ exhibits pronounced non-Arrhenius behavior as evident from the strong curvature in the Arrhenius representation of Fig. 2. It is common practice to use the Vogel-Fulcher-Tammann (VFT) function,

$$\tau_\alpha = \tau_0 \exp\left[\frac{DT_{VF}}{T-T_{VF}}\right] \qquad (3)$$

with $T_{VF}$ the Vogel-Fulcher temperature and $D$ the strength parameter [41] for the parameterization of such behavior. The solid line in Fig. 2 is a fit of $\tau_\alpha(T)$ at $T > T_g$ using this function. The resulting fit parameters are $\tau_0 = 4.2\times10^{-13}$ s, $T_{VF} = 189$ K, and $D = 3.8$. Within the strong-fragile classification scheme by Angell [41] the small value of $D$ characterizes BZP as fragile glass former. As an alternative quantitative measure of fragility, the parameter $m$ can be used [42] defined by the slope at $T_g$ in the Angell plot, $\log_{10}(\tau_\alpha)$ vs. $T_g/T$ (not shown). We obtain $m = 125$, corroborating the high fragility of BZP. As revealed in Fig. 2, the VFT function does not provide a perfect fit of the experimentally determined relaxation times. The deviations at sub-$T_g$ temperatures are due to the sample falling out of thermodynamic equilibrium when reaching the glass state. However also in equilibrium, at $T > T_g$, small but significant deviations show up, especially at high temperatures. Such a behavior is well known also for other glass-forming liquids and is commonly observed in experiments covering sufficiently broad frequency and temperature ranges [43].

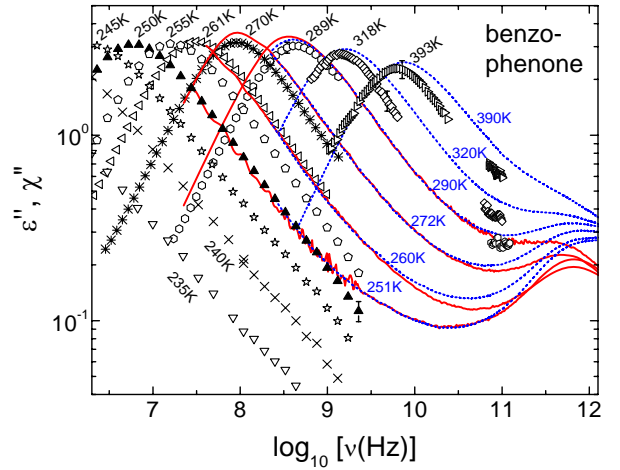

FIG. 3. (Color) Frequency dependence of the dielectric loss (symbols) and the imaginary part of the susceptibility obtained from OKE (blue dotted lines) [23] and DLS (red solid lines) [30] measurements. The susceptibility data were vertically scaled with a single factor for all temperatures to match the dielectric data at 250 K.

In Fig. 3, we plot our dielectric results at high temperatures together with the susceptibility spectra from OKE and DLS [23,30]. The $\alpha$ peak positions seem to be in rather good accord. In [23] relaxation times were reported, determined from fits of the OKE signal in time-domain. In addition, we extracted relaxation times from the peak positions $\nu_p$ of the DLS susceptibility [30] using $\tau_\alpha = 1/(2\pi\nu_p)$. Both data sets are shown in Fig. 2 (squares and

×'s). It is clear from earlier comparisons of light scattering and dielectric spectroscopy data that considerable deviations of the relaxation times determined by these methods can arise [2,44,45,46,47,48]. Usually $\tau$ from light scattering is smaller than that from dielectric spectroscopy. This can be understood considering different coupling to translational and reorientational degrees of freedom and different tensorial properties of the two methods [45,47]. In the present case (Fig. 2), there are small deviations of the DLS from the dielectric data, especially at high temperatures. This also becomes obvious if considering the peak positions in Fig. 3. However, these deviations may also be ascribable to small temperature errors due to imperfect thermal coupling and to the different methods of determining the relaxation times. Overall, the situation in BZP is similar as for glycerol where different methods lead to quite similar relaxation times [2,47,49]. It should be noted that there is also a difference concerning the widths of the OKE and DLS susceptibility peaks compared with those of the dielectric loss peaks (Fig. 3). The finding of somewhat smaller widths for the dielectric loss agrees with the reports in other materials [2,46].

In [23] the temperature dependence of $\tau_\alpha$ was analyzed using the critical law predicted by MCT in its simplest version, the so-called idealized MCT [4]:

$$\tau_\alpha \sim (T - T_c)^{-\gamma} \quad \text{with} \quad \gamma = \frac{1}{2a} + \frac{1}{2b} \qquad (4)$$

Good agreement with experimental data was stated and values of $T_c = 250$ K and $\gamma = 1.92$ were determined. The dashed line shown in Fig. 2 is a fit of the dielectric data at $T > 255$ K using these parameters. The much broader temperature and frequency range of the dielectric experiment reveals that eq. (4) is able to describe $\tau_\alpha(T)$ at high temperatures only. Also a free fit does not lead to better agreement. This behavior is a well established fact and commonly ascribed to thermally activated hopping processes, becoming important at temperatures close to and below $T_c$. It can be understood within extended versions of MCT only [4,50], whose application, however, is beyond the scope of the present work. Whatsoever, the inset of Fig. 2, corresponding to Fig. 4 of Ref. [23], demonstrates that critical behavior according to eq. (4) is well obeyed at high temperatures, $T > T_c$.

The width parameter $\beta_{CD}$ of the $\alpha$ relaxation resulting from the fits shown in Fig. 1 is provided in Fig. 4. Its values being clearly below unity signify non-Debye behavior, which is commonly ascribed to a distribution of relaxation times. $\beta_{CD}$ increases with increasing temperature and tends to saturate at a value below unity at high temperatures. This behavior is similar, e.g., to that found from broadband spectra of the molecular glass formers glycerol and propylene carbonate [2,51]. MCT predicts that for $T > T_c$, near $T_c$ the so-called time-temperature superposition principle should be obeyed, which implies a temperature-independent width parameter. In addition, non-Debye behavior at high temperatures, $T > T_c$ is expected. The experimental data in BZP (Fig. 4) are for the most part consistent with these predictions. However, as also found for glycerol and propylene carbonate, the observed saturation of $\beta_{CD}(T)$ occurs at a temperature somewhat higher than $T_c$ (reported values of $T_c$ in BZP are 250 K [23] and 235 K [25].

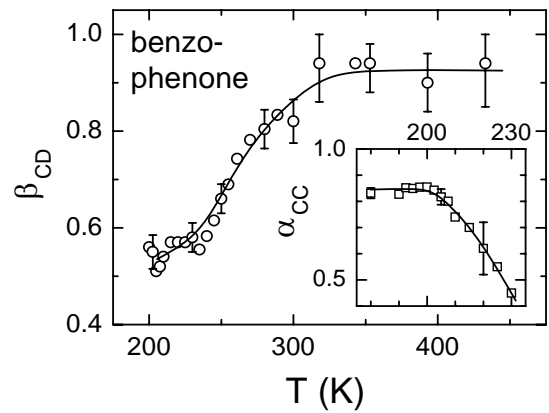

FIG. 4. Width parameters of the $\alpha$ (main frame) and the $\beta$ relaxation (inset). They were determined from the fits of the experimental loss spectra shown as solid lines in Fig. 1. The lines are drawn to guide the eyes.

When considering these results, one should be aware that in current literature there are some claims [32,52] that time-temperature superposition principle in glass formers may be valid even far below $T_c$ and that the commonly found temperature dependence of the width parameter is an artifact of the analysis. While in the present work a strong temperature dependence of $\beta_{CD}$ is obtained, an alternative type of evaluation applied to the OKE data of BZP [23] revealed a constant $\beta_{CD}$ of 0.71 [32]. However, the temperature range of 251 – 302 K considered in that analysis appears rather restricted if compared to that covered by Fig. 4. In this temperature region, our data suggest a variation of $\beta_{CD}$ by about 0.15 and, within the error bars, also would be consistent with a much weaker temperature dependence. One may speculate that the evaluation employed in Ref. [32], using CD fits that provide a good description in a limited region of the linear time-domain representation, may not reveal this variation. To check this question in more detail, in Fig. 5 a scaled plot of the dielectric loss, $\varepsilon''/\varepsilon_p$ vs. $\nu/\nu_p$, is provided, with $\varepsilon_p$ the peak height and $\nu_p$ the peak frequency. For constant width parameter, in this representation all $\alpha$ peaks should coincide. Already in the global plot of Fig. 5(a), it becomes obvious that the scaling fails at the high-frequency flanks of the $\alpha$-relaxation peaks, which is most clearly seen when restricting the plot to the $\alpha$ peak region (Fig. 5(b)). However, one should be aware that high-frequency processes as, e.g., secondary relaxations may have a non-negligible influence on the high-frequency flanks of the α peaks. In Fig. 5(a), for the two lowest temperatures shown, the β relaxation is seen. It has an amplitude that is much smaller than that of the α peak. Thus, judging from this figure and assuming a simple additive contribution, it seems unlikely that this relaxation could explain the failure of the α peak scaling. Also possible contributions from the fast β process seem unlikely to give rise to the observed temperature variation of the width parameter. If relevant at all, they should play the most important role for the highest temperature where the α peak approaches the frequency regime of the fast process. Thus, if correcting for these contributions, even higher values of $\beta_{CD}$ at high temperatures would result and its temperature dependence would be stronger than shown in Fig. 4. However, to finally clarify the issue of α-peak scaling and the temperature dependence of the width parameter, more additional high-precision data, extending to higher frequencies also for high temperatures and a more detailed analysis may be necessary.

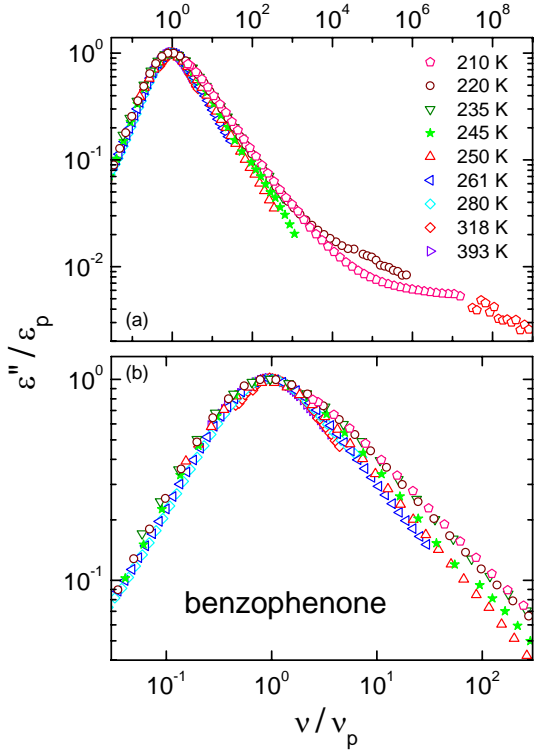

The relaxation strength $\Delta\varepsilon$ of BZP increases with decreasing temperature (Fig. 6). In the framework of the Onsager theory a Curie law, $\Delta\varepsilon \sim 1/T$, can be expected. However in the present case, similar to the findings, e.g., in glycerol [53,54], $\Delta\varepsilon(T)$ better can be parameterized by a Curie-Weiss-like temperature dependence, $\Delta\varepsilon \sim 1/(T-T_{CW})$, with a characteristic temperature of $T_{CW} = 107$ K (dashed line in Fig. 6). A stronger temperature dependence of $\Delta\varepsilon$ than Curie behavior is an often observed phenomenon in glass-forming materials [53,55] pointing to cooperative relaxation processes. As an alternative the experimental data can be analyzed within the framework of MCT. According to idealized MCT, at $T > T_c$ time-temperature superposition principle should be fulfilled, implying a temperature-independent relaxation strength. In addition, for $T < T_c$ a square-root singular behavior of the non-ergodicity parameter, which in the simplest case should be proportional to $\Delta\varepsilon$, follows from extended MCT, namely $f = c_1 + c_2 (T_c-T)^{1/2}$ ($c_1$ and $c_2$ are constants) [4]. This abrupt change of the non-ergodicity parameter at $T_c$ is often referred to as "cusp anomaly". It is interesting that if assuming $f \sim \Delta\varepsilon T$ instead of $f \sim \Delta\varepsilon$, the MCT predictions are nicely fulfilled for BZP (upper inset of Fig. 6). The resulting $T_c = 240$ K is close to the values reported in literature (250 K [23] and 235 K [25]). One may motivate the analysis of $\Delta\varepsilon T$ instead of $\Delta\varepsilon$ by a correction for the $1/T$ Curie law, which is not taken into account by MCT. This quantity also was considered in [55] for various glass formers but there disagreements with MCT were stated.

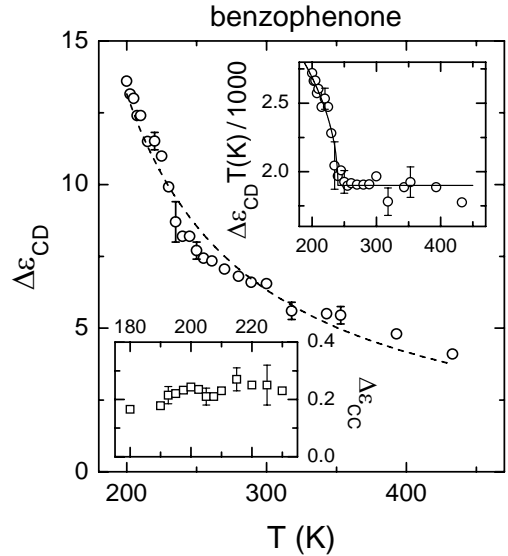

FIG. 5. (Color) Frequency dependence of the dielectric loss for selected temperatures, re-scaled to achieve identical peak frequencies and amplitudes of the alpha-relaxation peak. (a) shows the complete spectra, (b) the peak region.

FIG. 6. Relaxation strengths as obtained from the fits of the experimental loss spectra shown as solid lines in Fig. 1. The line in the main frame is a fit of the α relaxation strength using the Curie-Weiss law ($T_{CW} = 107$ K). In the upper inset $\Delta\varepsilon_{CD}$ multiplied by temperature is shown. The line is a fit with the MCT prediction (see text), leading to $T_c = 240$ K. The lower inset shows the relaxation strength of the β process.

## B. $\beta$ relaxation and excess wing

As already mentioned, our broadband spectra of BZP reveal a $\beta$ relaxation (Fig. 1). In most glass-forming materials, $\beta$ relaxations are mainly observed in the loss spectra at sub-$T_g$ temperatures where the $\alpha$ peak is shifted to very low frequencies and the $\beta$ relaxation leads to well-pronounced peaks [10,14,18,56]. Such a behavior is demonstrated for BZP in Fig. 7, which shows the spectra in the low-temperature region in more detail. Here very broad $\beta$ peaks show up in $\varepsilon''(\nu)$, continuously shifting to higher frequencies with increasing temperature. The spectra at $T \leq 197$ K were fitted using the CC function (dashed lines). The empirical CC function for most materials is found to be a good description of the experimental $\beta$ peaks, which also is the case for BZP. The strong increase at low frequencies observed for temperatures $T > 197$ K is due to the $\alpha$ relaxation peak. In Fig. 1 it is seen to continuously shift into the frequency window with increasing temperature. As mentioned in the preceding section, thus for the fits at 197 K < $T$ < 235 K the sum of a CD and a CC function was employed (solid lines in Fig. 1). With increasing temperature, the $\beta$ peak becomes successively submerged under the dominating $\alpha$ peak. In this way at 215 - 230 K an excess-wing arises and only the high-frequency measurements at $\nu > 1$ MHz provide the slight indication that it is not a simple second power law at the right flank of the $\alpha$ peak, but a shoulder. With increasing temperature the slope of the excess wing becomes steeper and finally only the $\alpha$ peak, with a single high-frequency flank, persists. The dashed lines in Fig. 1 show the CC part of the fits revealing a continuation of the trend seen in Fig. 7: The peak amplitude increases and the peak width decreases with temperature, both commonly observed features of $\beta$ relaxations.

The $\beta$ relaxation times $\tau_\beta = \tau_{CC}$ obtained from the fits are shown in Fig. 2 (open circles). At $T < T_g$, $\tau_\beta(T)$ can be fitted by a straight line, implying thermal activation with $E \approx 0.36$ eV. Such a behavior is often considered as a hallmark feature of $\beta$ relaxations. However, one should be aware that below $T_g$ even the $\alpha$ relaxation time usually follows an Arrhenius law (see, e.g., uppermost three points of $\tau_\alpha$ in Fig. 2). Reports of the $\beta$ relaxation time above the glass temperature are relatively scarce but in several glass formers it was found that at $T > T_g$, $\tau_\beta(T)$ exhibits much stronger non-Arrhenius temperature dependence [16,17,54,57,58,59,60,61]. As revealed by Fig. 2, such a behavior is also found for BZP. When considering these results, however, one should be aware that especially at elevated temperatures the determination of $\tau_\beta$ has a high uncertainty due to the strong overlap with the $\alpha$ process (see error bars in Fig. 2). In addition, one should mention that it is not so clear if a simple *additive* superposition of different contributions to $\varepsilon''(\nu)$ is really justified, especially if they strongly interfere with each other [62]. As an alternative we used the convolution approach promoted in [62], based on the product ansatz by Williams [63,64]. This analysis [65] revealed results, fully consistent with those shown in Fig. 2. As an example, the dotted lines in Fig. 1 represent fits of the 215 and 225 K spectra using this approach. Fits of equal quality as for the additive ansatz are achieved with identical $\alpha$ and $\beta$ relaxation times. Also in [66], by analyzing dielectric data on various materials, consistent results of the additive and the convolution ansatz were stated.

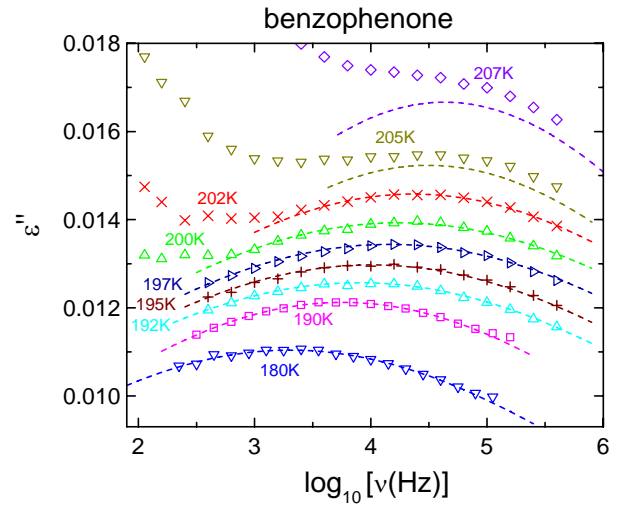

FIG. 7. (Color online) Frequency dependence of the dielectric loss at various temperatures below $T_g$. The dashed lines are fits with a CC function ($T < 197$ K) or the CC portion of the combined CC+CD fits shown in Fig. 1.

For a variety of glass formers, a correlation of the $\beta$ relaxation time at $T_g$ with the Kohlrausch exponent $\beta_{KWW}$ was found [67]. This was rationalized within the framework of the coupling model (CM) [5,6] where $\tau_\beta$ is expected to be approximately equal to the so-called primitive relaxation time $\tau_0$. $\tau_0$ can be calculated using the expression

$$\tau_0 = \left(t_c\right)^{1-\beta_{KWW}} \left(\tau_{KWW}\right)^{\beta_{KWW}} \qquad (5)$$

with $t_c$ of the order of 2 ps [6]. $\tau_{KWW}$ and $\beta_{KWW}$ are relaxation time and width parameter of the Kohlrausch-Williams-Watts (KWW) or stretched-exponential function [68] that often is used to describe the $\alpha$ relaxation. The relation $\tau_\beta \approx \tau_0$ recently even was proposed as a criterion to distinguish genuine JG $\beta$ relaxations, inherent to the glassy state of matter, from other types of secondary relaxations, e.g., due to intramolecular modes [69]. It should be noted that this criterion can be regarded as useful, even when doubting the validity of coupling model, as it is based on the purely phenomenological correlation mentioned above. Calculating the KWW parameters from the corresponding parameters of the CD function [70] (Figs. 2 and 4), we arrived at $\tau_0(T)$ as shown in Fig. 2 (pluses). At the higher temperatures, indeed good agreement of the prediction and the experimental data is achieved, while close to $T_g$ deviations of the order of ten are found. However, these deviations are within the range also found in other materials [17,60,69,71]. Overall, we can state that when adopting the criterion $\tau_\beta \approx \tau_0$ promoted in [69], the $\beta$ relaxation seen in BZP by dielectric spectroscopy seems to be of JG type.

The width parameter $\alpha_{CC}$ of the $\beta$ relaxation and its relaxation strength $\Delta\varepsilon_{CC}$ are provided in Figs. 4 (inset) and 5 (lower inset), respectively. While $\Delta\varepsilon_{CC}$ is only weakly temperature dependent, $\alpha_{CC}(T)$ of BZP decreases markedly at $T > T_g$. In the CC function, the quantities [28] $1-\alpha_{CC}$ and $\alpha_{CC}-1$ define the slopes of the left and right flank of the symmetric loss peak, respectively. $\alpha_{CC} = 0$ would imply monodispersive Debye behavior. Thus the detected decrease of $\alpha_{CC}$ mirrors the strong narrowing of the $\beta$ peak as seen in Fig. 1, which is typical for JG $\beta$ relaxations [14,56].

### C. Are MCT CC-peak and JG $\beta$ relaxation the same phenomena?

In the following we will address the question if the Cole-Cole peak at about 10 GHz, used in [24,25] to describe the OKE and DLS data on BZP, is a canonical $\beta$ relaxation peak [29]. As mentioned above, it is clearly revealed by Figs. 1 and 6 that BZP indeed has a canonical $\beta$ relaxation in the Hz-MHz range, most likely of JG type. However, with increasing temperature, after first developing into an excess wing, the $\beta$ peak finally completely merges with the $\alpha$ peak. Thus, already at 235 K, significantly below the lowest temperature investigated in the OKE and DLS experiments (251 K) [22,23,30] only a single peak is seen, without any trace of a shoulder or excess wing. Quite in contrast, at 251 K the most pronounced excess-wing like feature shows up in the OKE and DLS data (Fig. 3), which could be well described by the MCT CC peak [25]. The dielectric loss spectrum at this temperature is restricted to frequencies below 3 GHz. Nevertheless, as becomes obvious from Fig. 3, the frequency range is sufficient to reveal a significant deviation from the OKE and DLS spectra: It seems clear that the excess intensity in the OKE and DLS susceptibility at $\nu > 1$ GHz is not seen in $\varepsilon''(\nu)$. Thus, at first glance one could argue that the MCT CC peak, contributing to the high-frequency flank of the $\alpha$ peak in the OKE and DLS experiments, is not present in dielectric spectroscopy and thus not related in any way to the JG $\beta$ relaxation typically observed with this method.

However, this assumption is not conclusive. In early dielectric investigations aiming at the detection of the minimum caused by the fast process of MCT, which were restricted to frequencies of the order of 10 GHz, its absence was stated [13,55,72]. In the early days of MCT, this lead to the speculation that the theory may be invalid. However, subsequent measurements in a broader frequency range indeed revealed such a minimum [73,74] and it soon became clear that its amplitude, relative to that of the $\alpha$ peak, usually is much weaker in dielectric spectra than in those obtained by scattering methods [2,46,74,75]. In the meantime a variety of extensions of MCT have been formulated demonstrating that the mentioned differences can be understood taking into account different coupling to density fluctuations and different tensorial properties of the experimental probes [45,75,76]. The MCT CC peak found in Refs. [24,25] arises from the same fast $\beta$ dynamics as the minimum of the original MCT and thus in dielectric spectra it also should have smaller amplitude than in scattering experiments. This could easily explain the differences observed in the 251 K spectra at $\nu > 1$ GHz and the fact that the $\beta$ relaxation in $\varepsilon''(\nu)$ is already merged with the $\alpha$ relaxation at 235 K while it still may contribute to the OKE and DLS data at much higher temperatures. However, in this context it should be noted that some support for the notion of a different origin of MCT CC peak and slow $\beta$ relaxation arises from a recent work comparing the manifestation of the slow $\beta$ relaxation in light scattering and dielectric spectra of various glass formers [77]. There much smaller amplitudes of this feature in light scattering were found, which is in contrast to the experimentally detected and theoretically predicted larger amplitude of the fast MCT process.

Further dielectric measurements in the range 250 K – 300 K, extending to frequencies comparable to those of the scattering experiments are needed to clarify if here the CC peak of MCT also contributes to the dielectric spectra. In Figs. 1 and 3, first results from quasi-optic measurements at 60 – 120 GHz are shown, which, however, are too incomplete to allow for definite conclusions. At least it seems to become clear from these data that also dielectric loss spectra of BZP may exhibit a high-frequency minimum. As revealed by the results at about 290 K, it seems to roughly agree with the minimum observed in the OKE and DLS spectra. Due to experimental problems (section II), currently we cannot provide more complete spectra in this region and a thorough high-frequency characterization of BZP has to be postponed to a forthcoming paper.

While the present spectra themselves do not a allow for a definite answer of the title question of this section, the relaxation times determined from these spectra (Fig. 2) may give some further hints. In [25], from the MCT analysis of the OKE spectra the frequency of the CC peak was determined as 10.7 GHz. The corresponding relaxation time of 14.9 ps is indicated by the single star in Fig. 2. Judging from this Figure, it seems very unlikely that the $\tau_\beta(T)$ curve obtained from the analysis of the dielectric data (open circles) should match the value from the MCT analysis at 251 K. In fact, according to the current status of the theory, the relaxation time of the CC peak predicted by MCT should be temperature independent [25]. The temperature dependence of the OKE data of Ref. [25] could be described by the variation of the $\alpha$ peak alone, its high-frequency flank (the von Schweidler law) strongly superimposing the temperature independent CC peak. However, the broader temperature range of the dielectric experiment reveals a variation of the $\beta$ peak, which is too strong to be taken into account in this way.

In this context it is of interest that eqs. (16a) and (16b) of Ref. [25] predict a decrease of the CC relaxation time with increasing CC peak amplitude. As mentioned above, it is well known that the amplitude of the fast $\beta$ process in susceptibility spectra depends on the experimental technique used. Judging from earlier investigations comparing the response from light scattering and dielectric spectroscopy at high frequencies [2,45,46,74,75] the fast process detected in dielectric spectra has an amplitude that can be up to one order of magnitude weaker. The CC peak is a manifestation of the fast $\beta$ relaxation and thus its dielectric amplitude also may be weaker. Making the reasonable assumption that all other parameters determining the CC peak frequency [25] do not depend on experimental probe, therefore for dielectric spectroscopy on BZP the relaxation time of the MCT CC peak may be expected at an up to one decade higher value than the star shown in Fig. 2. However, even then an extrapolation of the experimentally detected points (open circles) to such a value seems difficult.

A final point to be considered in this section is the temperature dependence and absolute value of the width parameter $\alpha_{CC}$ as shown in the inset of Fig. 4. As mentioned in the preceding section, the quantities $1-\alpha_{CC}$ and $\alpha_{CC}-1$ are the exponents of the limiting power laws of $\varepsilon''(v)$ for frequencies much smaller and much larger than the CC peak frequency, respectively. In the MCT framework outlined in [24,25], they are identified with the parameter $a$ (or $-a$, respectively). This so-called critical exponent of MCT is temperature independent [4]. However, the experimental data shown in the inset of Fig. 4 reveal strong temperature dependence, a behavior commonly observed for slow $\beta$ relaxations. In addition, at $T > 220$ K, $\alpha_{CC}$ seems to reach values below 0.6. This would imply $a > 0.4$, which is outside of the allowed range of values for this quantity within MCT [3,4].

## IV. SUMMARY AND CONCLUSIONS

In the present work we have provided a thorough characterization of the relaxational behavior of glass forming BZP. With a relaxation time following VFT behavior (Fig. 2), a width parameter $\beta$ increasing (Fig. 4) and a relaxation strength decreasing with increasing temperature (Fig. 6), the $\alpha$ relaxation behaves rather conventional. The temperature dependence of the $\alpha$ relaxation time characterizes BZP as a fragile glass former. In most respects, the behavior of the $\alpha$ relaxation is in good accord with MCT; in fact we know of no other glass formers where the predictions are so well fulfilled by the dielectric spectroscopy results: The relaxation time shows the expected critical behavior at $T > T_c$, in good accord with the results of the earlier OKE investigation [23] (inset of Fig. 2). The relaxation strength exhibits the predicted cusp anomaly (upper inset of Fig, 5). Finally, the width parameter saturates to a value below unity at high temperatures, even though this value is reached somewhat above $T_c$ only (Fig. 4).

Strong interest in this material arises from the fact that the OKE spectra in the 200 MHz – 5 THz region could be described within MCT by a CC peak [24,25], which reminds of a slow $\beta$ relaxation. As revealed by our broadband measurements, at low temperatures BZP indeed has a well pronounced slow $\beta$ relaxation, most likely of JG type. At higher temperatures, it develops into an excess wing and finally merges with the $\alpha$ peak. We now will list the facts that are relevant to clarify the relation of this canonical $\beta$ relaxation to the MCT CC peak: i) The $\beta$ peak becomes submerged under the $\alpha$ peak already at 235 K (Fig. 1), while the excess-wing like feature described by the MCT CC peak in the OKE data is best seen at 251 K. ii) Comparing the dielectric and OKE spectra up to 2 GHz at this temperature, they do not agree and $\varepsilon''(v)$ does not show any sign of an excess wing (Fig. 3). iii) The relaxation time from the MCT analysis of the OKE data at 251 K [25] does not match the extrapolated $\tau_\beta(T)$ from dielectric spectroscopy. iv) The frequency and width parameter of the $\beta$ peak in BZP are strongly temperature dependent, while the MCT CC peak should be temperature independent. In addition, the width parameter reaches values outside the range allowed within MCT.

Overall, it is clear from these findings that the excess-wing-like feature seen by OKE/DLS and analyzed by the MCT CC-peak cannot be a simple continuation of the slow $\beta$ peak seen in the dielectric spectra at lower temperatures and frequencies. A comparison of dielectric and light scattering spectra at high frequencies in other glass formers revealed a weaker amplitude of the fast MCT process in the dielectric case [2,46,74,75]. This could explain the merging of the $\beta$ relaxation in the dielectric spectra, already at 235K. However, it seems unlikely that the weaker coupling to the fast process for the dielectric probe also could explain the strong discrepancy in the relaxation times (Fig. 2) in terms of the amplitude dependence of the CC relaxation time [25]. In any case, our results never could be described with a temperature-independent CC-peak, which only is possible for the much smaller temperature and frequency range of the OKE experiment [22,23,24,25]. We want to point out that these findings do not imply that the description of the OKE data with a CC peak arising from the fast $\beta$ relaxation of MCT is invalid. In addition, we cannot exclude that this peak also is present in the dielectric loss, even though it should be of weaker amplitude to explain the discrepancy of the 251 K spectra. However, according to the current status of the theory, this CC peak seems to be a separate phenomenon and not related to the JG $\beta$ relaxation as detected by dielectric spectroscopy in BZP. To achieve the final goal of covering all dynamic processes including the JG $\beta$ relaxation, future theoretical developments would need to reveal a strong temperature dependence of both the CC-peak frequency and width.

## ACKNOWLEDGMENTS

We thank M. Sperl for stimulating discussions and helpful comments and E.A. Rössler for providing the files of the DLS and OKE data shown in Fig. 3. One of the authors (L.C.P.) would like to acknowledge the financial support of the Humboldt Foundation, Ministerio de Educación y Ciencia (Grants No. FIS2005-00975 and No. MEC-EX2005), and of Generalitat de Catalunya (Grant No. SGR2005-00535).